\documentclass[prb,amsfonts,twocolumn,showpacs,amssymb,floatfix,bibnotes]{revtex4}
\usepackage{bm}
\usepackage{tabularx}
\usepackage{ascmac}
\usepackage{amsmath}
\usepackage{wrapfig}
\usepackage{graphicx}
\usepackage{float}
\usepackage[FIGBOTCAP]{subfigure}
\usepackage{color}

\bmdefine{\bolds}{s}
\bmdefine{\boldi}{i}
\bmdefine{\boldj}{j}
\bmdefine{\boldtau}{\tau}
\bmdefine{\boldsigma}{\sigma}
\bmdefine{\boldlambda}{\lambda}
\bmdefine{\boldk}{k}
\bmdefine{\boldK}{K}
\bmdefine{\boldq}{q}
\bmdefine{\boldQ}{Q}
\bmdefine{\boldr}{r}
\bmdefine{\boldj}{j}

\begin{document}


\title{
Competition between spin fluctuations 
in Ca$_{2-x}$Sr$_{x}$RuO$_{4}$ around $x=0.5$
}


\author{Naoya Arakawa}
\email{arakawa@hosi.phys.s.u-tokyo.ac.jp} 
\author{Masao Ogata}
\affiliation{Department of Physics, 
The University of Tokyo,
Tokyo 113-0033, Japan}


\date{\today}

\begin{abstract}
We study the static susceptibilities for charge and spin sectors 
in paramagnetic states for Ca$_{2-x}$Sr$_{x}$RuO$_{4}$ in $0.5\leq x \leq 2$ 
within random phase approximation 
on the basis of an effective Ru $t_{2g}$ orbital Hubbard model. 
We find that 
several modes of spin fluctuation around $\boldq=(0,0)$ and 
$\boldq\sim(0.797\pi,0)$ 
are strongly enhanced for the model of $x=0.5$. 
This enhancement arises from 
the increase of the corresponding susceptibilities 
for the $d_{xy}$ orbital due to 
the rotation-induced modifications of the electronic structure 
for this orbital (i.e., the flattening of the bandwidth 
and the increase of the density of states near the Fermi level). 
We also find that 
the ferromagnetic spin fluctuation becomes stronger 
for a special model than for the model of $x=0.5$, 
while the competition 
between the modes of spin fluctuation at $\boldq=(0,0)$ and 
around $\boldq\sim (\pi,0)$ is weaker for the special model; 
in this special model, 
the van Hove singularity (vHs) for the $d_{xy}$ orbital 
is located on the Fermi level. 
These results indicate that 
the location of the vHs for the $d_{xy}$ orbital, 
which is controlled by substitution of Ca for Sr, 
is a parameter to control this competition. 
We propose that 
the spin fluctuations for the $d_{xy}$ orbital 
around $\boldq=(0,0)$ and $\boldq\sim (\pi,0)$ 
play an important role in the electronic states around $x=0.5$ 
other than the criticality approaching the usual Mott transition 
where all electrons are localized. 
\end{abstract}


\maketitle

\section{Introduction}
Ca$_{2-x}$Sr$_{x}$RuO$_{4}$ is one of the strongly correlated electron systems 
with orbital degrees of freedom.~\cite{Nakatsuji-discovery,Nakatsuji-lattice} 
This alloy has a quasi-two-dimensional crystalline structure, 
and the Ru $t_{2g}$ orbitals and the lattice distortions 
induced by substitution of Ca for Sr 
play important roles in determining the electronic structures. 
Actually, this alloy shows 
a spin-triplet superconductivity,~\cite{Maeno-triplet,Ishida-Knight-shift,
neutron-triplet} 
a heavy fermion (HF) behavior,~\cite{Nakatsuji-HF} 
and a metal-insulator transition,~\cite{Nakatsuji-MI} 
depending on the Sr concentration. 

To clarify the origin of the HF behavior around $x=0.5$, 
the present authors studied 
the electronic states for Ca$_{2-x}$Sr$_{x}$RuO$_{4}$ in $0.5\leq x \leq 2$ 
within the Gutzwiller approximation 
on the basis of an effective Ru $t_{2g}$ orbital Hubbard model;~\cite{NA-GA} 
this effective model takes account of 
the change of the $dp$ hybridizations 
due to the rotation of RuO$_{6}$ octahedra, 
which is induced in the range of $x < 1.5$. 
We proposed that 
this HF behavior can be qualitatively understood as 
the cooperative effect between moderately strong electron correlation 
and the orbital-dependent modification of the electronic structures 
due to the rotation of RuO$_{6}$ octahedra.~\cite{NA-GA} 
In that study, however, 
we considered only the local correlations. 
It is thus needed to study the effects of the nonlocal correlation 
on the electronic structures, 
since both local and nonlocal correlations play important roles 
in discussing the electronic structure for a strongly correlated electron system 
in general. 

In this paper, 
we study the static susceptibilities for charge and spin sectors 
in paramagnetic (PM) states for Ca$_{2-x}$Sr$_{x}$RuO$_{4}$ in $0.5\leq x \leq 2$ 
within random phase approximation (RPA) 
on the basis of the effective Ru $t_{2g}$ orbital Hubbard model.~\cite{NA-GA} 
In particular, we analyze these static susceptibilities 
for the models of $x=2$ and $0.5$, and a special model 
to clarify effects of the rotation of RuO$_{6}$ octahedra, 
the van Hove singularity (vHs) for the $d_{xy}$ orbital, 
and the Hund's rule coupling. 
In this special model, 
the vHs for the $d_{xy}$ orbital is located on the Fermi level. 

The paper is organized as follows. 
In Sec. II, we first show the effective Ru $t_{2g}$ orbital Hubbard model 
for Ca$_{2-x}$Sr$_{x}$RuO$_{4}$ 
in both the absence and the presence of the rotation of RuO$_{6}$ octahedra. 
Next, we introduce a generalized susceptibility 
for a multiorbital system 
and susceptibilities for charge and spin sectors. 
Then, we explain the method to determine a primary instability 
for a multiorbital system. 
In Sec. III, 
we first show the static susceptibilities 
and the maximum eigenvalues of static susceptibility 
for the models of $x=2$ and $0.5$ 
to analyze the effects of the rotation of RuO$_{6}$ octahedra. 
We next compare the results for the model of $x=0.5$ 
with those for the special model 
to analyze the effects of the vHs for the $d_{xy}$ orbital. 
Then, we show the dependence of the static susceptibilities 
on the Hund's rule coupling 
for the models of $x=2$ and $0.5$, and the special model. 
In Sec. IV, we first address the mass enhancement 
due to the enhanced fluctuations for the models of $x=2$ and $0.5$. 
In addition, 
we compare our results with previous theoretical 
and experimental works about magnetic properties. 
The paper concludes with a summary of our results in Sec. V. 

\section{Formulation}
In the following, 
we set $\hbar=\mu_{\textrm{B}}=k_{\textrm{B}}=1$, 
label the $d_{xz}$, $d_{yz}$, and $d_{xy}$ orbitals as $1$, $2$, and $3$, 
and choose the coordinates, $x$, $y$, and $z$, 
in the directions of the Ru-O bonds 
at $\phi=$ 0$^{\circ}$ (i.e., $1.5\leq x \leq 2$). 
Here, $\phi$ is an angle of the rotation of RuO$_{6}$ octahedra. 

\begin{figure}[tb]
{\includegraphics[width=80mm]{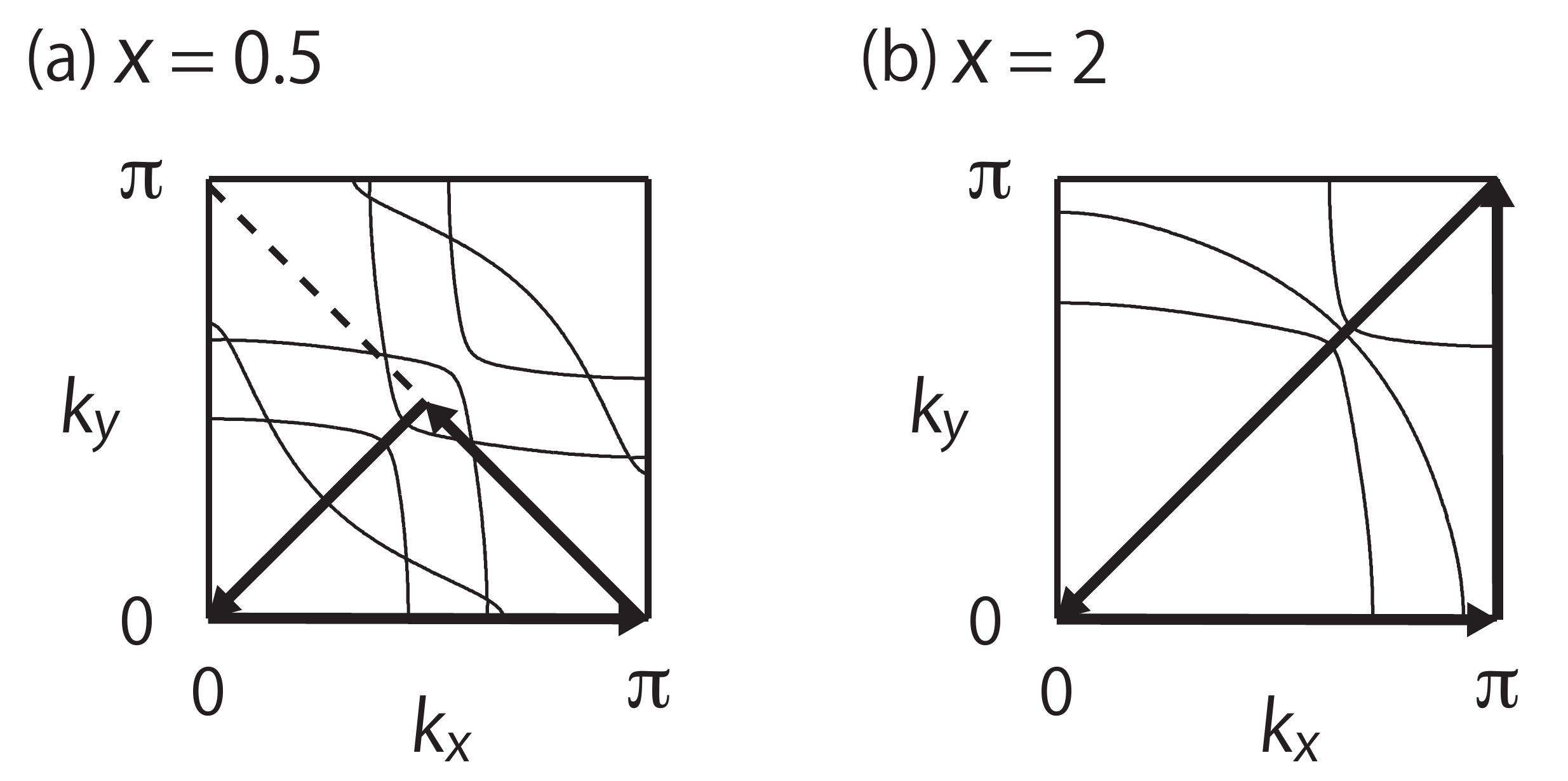}}
\vspace{-14pt}
\caption{FSs for the models of (a) $x=0.5$ and (b) $2$. 
The dashed line in Fig. (a) represents the folded BZ. 
The arrows in Figs. (a) and (b) correspond to 
the momentum lines used in the following figures. }
\label{fig:FS}
\end{figure}

To study the electronic states 
for Ca$_{2-x}$Sr$_{x}$RuO$_{4}$ in $0.5\leq x \leq 2$, 
we consider an effective Ru $t_{2g}$ orbital Hubbard model 
which takes account of the change of the $dp$ hybridizations 
between the Ru $4d$ and O $2p$ orbitals 
due to the rotation of RuO$_{6}$ octahedra~\cite{NA-GA}: 
\begin{align}
\hat{H}=\hat{H}_{0}+\hat{H}_{\textrm{int}} . \label{eq:Hubbard}
\end{align}
In the presence of the rotation of RuO$_{6}$ octahedra, 
the alternation of the direction of the rotation 
in the two-dimensional square lattice leads to a unit cell doubled, 
and the Brillouin zone (BZ) is folded [see Fig. \ref{fig:FS} (a)]. 

The noninteracting Hamiltonian 
in the absence of the rotation of RuO$_{6}$ octahedra is given by
\begin{align}
\hat{H}_{0}
=
\textstyle\sum\limits_{\boldk}
\textstyle\sum\limits_{a,b=1}^{3}
\textstyle\sum\limits_{s=\uparrow,\downarrow}
\epsilon_{ab}(\boldk, 0^{\circ})
\hat{c}^{\dagger}_{\boldk a s} 
\hat{c}_{\boldk b s} , \label{eq:H0-absence}
\end{align}
where $\epsilon_{ab}(\boldk, 0^{\circ})$ 
denotes the energy dispersions for $\phi=0^{\circ}$. 
These energy dispersions measuring from the chemical potential, $\mu$, 
are given by
\begin{align}
\epsilon_{11}(\boldk, 0^{\circ})&
=-2 t_{1} \cos k_{x}
-2 t_{2} \cos k_{y} -\mu, \label{eq:dis11-0}\\
\epsilon_{12}(\boldk, 0^{\circ})&
=\ \epsilon_{21}(\boldk, 0^{\circ}) 
= -4 t^{\prime} \sin k_{x} \sin k_{y}, \label{eq:dis12-0}\\
\epsilon_{22}(\boldk, 0^{\circ})&
=-2t_{2} \cos k_{x}
-2t_{1} \cos k_{y} -\mu, \label{eq:dis22-0}\\
\epsilon_{33}(\boldk, 0^{\circ})&
=-2t_{3}(\cos k_{x}+\cos k_{y})
-4t_{4}\cos k_{x} \cos k_{y} -\mu, 
\label{eq:dis33-0}\\
\epsilon_{ab}(\boldk, 0^{\circ})&
=\ 0 \ \ \ \ \textrm{otherwise}\ . \label{eq:disab-0} 
\end{align} 
In this study, we take account of 
the next-nearest-neighbor (NNN) hopping integral 
between the $d_{xz}$ and $d_{yz}$ orbitals, 
which has been neglected in the previous study (Ref.\onlinecite{NA-GA}). 
$\mu$ is determined so that 
the total occupation number for the Ru $t_{2g}$ orbitals is equal to 4. 

For the model of $x=2$, 
we set $t_{1}=0.675$, $t_{2}=0.09$, $t_{3}=0.45$, $t_{4}=0.18$, 
and $t^{\prime}=0.03$ so as to reproduce 
the experimentally observed Fermi surfaces (FSs).~\cite{dHvA} 
Here and throughout this paper, the energy unit is eV. 
Figure \ref{fig:FS} (b) shows the FSs for this model. 
We see that 
the NNN hopping integral between the $d_{xz}$ and $d_{yz}$ orbitals 
leads to the slight change of the topology of the FS 
around $\boldk\sim (0.67\pi,0.67\pi)$ from that 
without this NNN hopping integral (Ref.\onlinecite{NA-GA}); 
there are no qualitative changes in the density of states (DOS), 
which is not shown here. 

In the presence of the rotation of the RuO$_{6}$ octahedra, 
the noninteracting Hamiltonian becomes 
\begin{align}
\hat{H}_{0}
=
\textstyle\sum\limits_{\boldk}^{\prime}
\textstyle\sum\limits_{a,b=1}^{3}
\textstyle\sum\limits_{l,l^{\prime}=A,B}
\textstyle\sum\limits_{s=\uparrow,\downarrow}
\epsilon_{ab}^{l l^{\prime}}(\boldk, \phi)
\hat{c}^{\dagger}_{\boldk a l s} 
\hat{c}_{\boldk b l^{\prime} s} ,\label{eq:H0-presence}
\end{align}
where the prime in the summation with respect to momentum 
represents the restriction within the folded BZ for $\phi\neq 0^{\circ}$, and 
$l$ and $\epsilon_{ab}^{l l^{\prime}}(\boldk, \phi)$ denote the sublattice index 
and the energy dispersions for $\phi \neq 0^{\circ}$. 
These energy dispersions measuring from $\mu$ are given by 
\begin{align}
\epsilon_{11}^{A A}(\boldk, \phi)
=& \ 
\epsilon_{22}^{A A}(\boldk, \phi)
= \dfrac{1}{3}\Delta_{t_{2g}}-\mu,\label{eq:dis11AA-phi} \\ 
\epsilon_{12}^{A A}(\boldk, \phi)
=&\ 
\epsilon_{21}^{A A}(\boldk, \phi)
= -4 t^{\prime} \sin k_{x} \sin k_{y}, \label{eq:dis12AA-phi}\\
\epsilon_{33}^{A A}(\boldk, \phi)
=& -\dfrac{2}{3}\Delta_{t_{2g}}-4t_{4}\cos k_{x} \cos k_{y}
-\mu,\label{eq:dis33AA-phi} \\
\epsilon_{11}^{A B}(\boldk, \phi)
=& 
-2t_{1} \cos^{2} \phi \cos k_{x}
-2(t_{2} -t_{1}\sin^{2} \phi) \cos k_{y}, 
\label{eq:dis11AB-phi}\\
\epsilon_{12}^{A B}(\boldk, \phi)
=&\ t_{1} \sin 2\phi ( \cos k_{x} + \cos k_{y}), 
\label{eq:dis12AB-phi}\\
\epsilon_{21}^{A B}(\boldk, \phi)
=& - \epsilon_{12}^{A B}(\boldk, \phi), \label{eq:dis21AB-phi}\\
\epsilon_{22}^{A B}(\boldk, \phi)
=&  
-2(t_{2} -t_{1}\sin^{2} \phi) \cos k_{x}
-2t_{1} \cos^{2} \phi \cos k_{y}, 
\label{eq:dis22AB-phi}\\
\epsilon_{33}^{A B}(\boldk, \phi)
=& 
-2t_{3} \cos^{3} 2\phi ( \cos k_{x} + \cos k_{y})\notag\\
&+2t_{5} \cos 2\phi \sin^{2} 2\phi
( \cos k_{x} + \cos k_{y})\notag\\
&-4t_{6} \cos 2\phi \sin^{2} 2\phi
( \cos k_{x} - \cos k_{y}), 
\label{eq:dis33AB-phi}\\
\epsilon_{ab}^{A A}(\boldk, \phi)
=& \ 
\epsilon_{ab}^{A B}(\boldk, \phi) 
= 0 
\ \ \ \ \ \ \textrm{otherwise} ,\label{eq:disab-phi}\\ 
\epsilon_{ab}^{B B}(\boldk, \phi) 
=& \ \epsilon_{ab}^{A A}(\boldk, -\phi),\label{eq:disab-BA-1}\\
\epsilon_{ab}^{B A}(\boldk, \phi) 
=& \ \epsilon_{ab}^{A B}(\boldk, -\phi).\label{eq:disab-BA-2}
\end{align}
The detail of the derivation of these energy dispersions 
is described in Ref.~\onlinecite{NA-GA}. 
In addition to the change of the hopping integrals 
due to the rotation of RuO$_{6}$ octahedra, 
we have taken account of the effect of the rotation-induced hybridization 
of the $d_{xy}$ orbital to the $d_{x^{2}\textrm{-}y^{2}}$ orbital 
as the difference of the crystalline-electric-field (CEF) energies 
between the $d_{xz/yz}$ and $d_{xy}$ orbitals, $\Delta_{t_{2g}}$.~\cite{NA-GA} 

\begin{figure}[tb]
\vspace{-4pt}
\includegraphics[width=80mm]{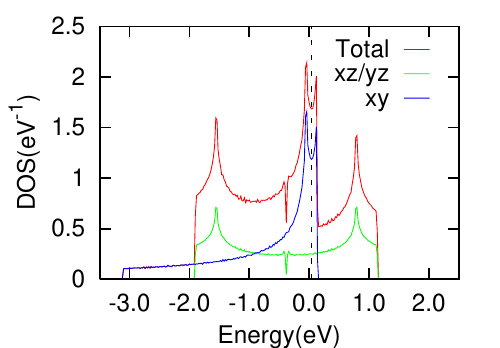}
\vspace{-10pt}
\caption{(Color online) DOS for the model of $x=0.5$. 
The dashed black line represents 
the chemical potential. 
}
\label{fig:DOS-x05}
\end{figure}

For the model of $x=0.5$, 
we set $\phi=15^{\circ}$, $\Delta_{t_{2g}}=0.39$, and $t_{5}=t_{6}=0$ 
so as to reproduce the experimentally observed FSs.~\cite{ARPES05} 
Note that the value of $\Delta_{t_{2g}}$ in this study 
is different from that in the previous studies~\cite{NA-GA,NA-ICM} 
due to the introduction of the NNN hopping integral 
between the $d_{xz}$ and $d_{yz}$ orbitals. 
Figures \ref{fig:FS} (a) and \ref{fig:DOS-x05} 
show the FSs and DOS for this model. 
We see from Fig. \ref{fig:DOS-x05} that 
the n.n.n. hopping integral between the $d_{xz}$ and $d_{yz}$ orbitals 
leads to a sharp decrease of the DOS for the $d_{xz}$ and $d_{yz}$ orbitals 
around an energy of $-0.4$ 
due to a small gap opening in the bands for these orbitals. 
This result is consistent with that obtained 
in the density-functional calculation for $x=0.5$.~\cite{Oguchi-LS} 

In addition, 
to discuss the role of the vHs for the $d_{xy}$ orbital, 
we consider the special model, 
for which we set $\phi=15^{\circ}$, $\Delta_{t_{2g}}=0.27$, and $t_{5}=t_{6}=0$; 
similarly to the model of $x=0.5$, 
the value of $\Delta_{t_{2g}}$ is different from that 
in the previous studies~\cite{NA-GA,NA-ICM}. 
In this special model, 
the vHs for the $d_{xy}$ orbital is located on the Fermi level, 
as denoted in Sec. I. 

We assume that 
the interacting Hamiltonian is given by the standard on-site interactions: 
\begin{align}
\hat{H}_{\textrm{int}} 
=&\ U 
\textstyle\sum\limits_{\boldj}
\textstyle\sum\limits_{a}
\hat{n}_{\boldj a \uparrow} \hat{n}_{\boldj a \downarrow}
+ U^{\prime}  
\textstyle\sum\limits_{\boldj}
\textstyle\sum\limits_{a>b}
\hat{n}_{\boldj a} \hat{n}_{\boldj b}\notag\\
&- 
J_{\textrm{H}} 
\textstyle\sum\limits_{\boldj}
\textstyle\sum\limits_{a>b}
( 
2 \hat{\bolds}_{\boldj a} \cdot 
\hat{\bolds}_{\boldj b} 
+ 
\frac{1}{2} \hat{n}_{\boldj a} \hat{n}_{\boldj b} 
)\notag\\
&+
J^{\prime} 
\textstyle\sum\limits_{\boldj}
\textstyle\sum\limits_{a>b}
\hat{c}_{\boldj a \uparrow}^{\dagger} 
\hat{c}_{\boldj a \downarrow}^{\dagger} 
\hat{c}_{\boldj b \downarrow} 
\hat{c}_{\boldj b \uparrow},\label{eq:Hint}
\end{align}
where 
$\hat{n}_{\boldj a}=\sum_{s}\hat{n}_{\boldj a s}=\sum_{s,s^{\prime}}
\hat{c}^{\dagger}_{\boldj a s}\sigma_{s s^{\prime}}^{0}\hat{c}_{\boldj a s^{\prime}}$ 
and $\hat{\bolds}_{\boldj a}=\frac{1}{2}
\sum_{s,s^{\prime}}\hat{c}^{\dagger}_{\boldj a s} 
\boldsigma_{s s^{\prime}} \hat{c}_{\boldj a s^{\prime}}$ 
with $\sigma_{s s^{\prime}}^{0}$ and $\boldsigma_{s s^{\prime}}$ 
being the unit matrix and the Pauli matrices.

To analyze fluctuations for charge and spin sectors 
for a multiorbital system, 
we define a generalized susceptibility at temperature $T$ as 
\begin{align}
&\chi_{a s_{1}, b s_{2}; c s_{3}, d s_{4}}^{l l^{\prime}}(\boldq,\tau)\notag\\
&=
\textstyle\sum\limits_{\boldk,\boldk^{\prime}}^{\prime}
\langle \textrm{T}_{\tau}  
\hat{c}_{\boldk^{\prime} a l s_{1}}^{\dagger}(\tau)
\hat{c}_{\boldk^{\prime}+\boldq b l s_{2}}(\tau)
\hat{c}_{\boldk c l^{\prime} s_{3}}^{\dagger}
\hat{c}_{\boldk-\boldq d l^{\prime} s_{4}} \rangle ,
\label{eq:generalized-chi}
\end{align}
and the corresponding Fourier coefficient is 
\begin{align}
&\chi_{a s_{1}, b s_{2}; c s_{3}, d s_{4}}^{l l^{\prime}}(q)
=\int^{T^{-1}}_{0}
\hspace{-14pt}
d\tau e^{i\tau \Omega_{n}}
\chi_{a s_{1}, b s_{2}; c s_{3}, d s_{4}}^{l l^{\prime}}(\boldq,\tau) . 
\end{align}
Here, $\textrm{T}_{\tau}$ is 
a chronological operator for imaginary time $\tau$, 
$\hat{c}_{\boldk a l \sigma}(\tau) 
= e^{\tau\hat{H}}\hat{c}_{\boldk a l \sigma}e^{-\tau\hat{H}}$, 
and $q\equiv (\boldq,i \Omega_{n})$ with 
$\Omega_{n}$ being a bosonic Matsubara frequency. 
In a PM state, the generalized susceptibility satisfies 
\begin{align}
\chi_{a\downarrow,b\downarrow;c\downarrow,d\downarrow}^{l l^{\prime}}(q)
=&\ \chi_{a\uparrow,b\uparrow;c\uparrow,d\uparrow}^{l l^{\prime}}(q),\\ 
\chi_{a\downarrow,b\uparrow;c\uparrow,d\downarrow}^{l l^{\prime}}(q)
=&\ \chi_{a\uparrow,b\downarrow;c\downarrow,d\uparrow}^{l l^{\prime}}(q).
\end{align}

The noninteracting susceptibility is 
\begin{align}
\chi_{abcd}^{l l^{\prime} (0)}(q)
=-\dfrac{1}{TN} 
\textstyle\sum\limits_{\boldk}^{\prime} 
\textstyle\sum\limits_{m} 
G_{da}^{l^{\prime} l (0)}(k+q)G_{bc}^{l l^{\prime} (0)}(k).
\end{align}
Here, $k\equiv (\boldk,i\omega_{m})$ with 
$\omega_{m}$ being a fermionic Matsubara frequency, and 
$G_{ab}^{l l^{\prime} (0)}(k)$ is the noninteracting Green's function: 
\begin{align}
G_{ab}^{l l^{\prime} (0)}(k)
= 
\textstyle\sum\limits_{\alpha}
(U_{\boldk})_{al;\alpha}
\dfrac{1}{i\omega_{m}-\epsilon_{\alpha}(\boldk,\phi)}
(U_{\boldk}^{\dagger})_{\alpha;bl^{\prime}}, 
\end{align}
where $(U_{\boldk})_{al;\alpha}$ is a unitary matrix 
to diagonalize $\hat{H}_{0}$: 
$\hat{c}_{\boldk a l s}=\sum_{\alpha} 
(U_{\boldk})_{al;\alpha}\hat{c}_{\boldk \alpha s}$ and 
\begin{align}
\epsilon_{\alpha}(\boldk,\phi)
=\textstyle\sum\limits_{a,b=1}^{3}
\textstyle\sum\limits_{l,l^{\prime}=A,B}
(U_{\boldk}^{\dagger})_{\alpha;al}\epsilon_{ab}^{l l^{\prime}}(\boldk,\phi)
(U_{\boldk})_{bl^{\prime};\alpha}.
\end{align} 

Let us introduce susceptibilities for charge and spin sectors as follows: 
\begin{align}
\chi_{abcd}^{l l^{\prime} (\textrm{C})}(q)
=& \ 
\chi_{a \uparrow, b \uparrow; c \uparrow, d \uparrow}^{l l^{\prime}}(q)
+
\chi_{a \downarrow, b \downarrow; c \uparrow, d \uparrow}^{l l^{\prime}}(q),\label{eq:chiC-def}\\
\chi_{abcd}^{l l^{\prime} (\textrm{S})}(q)
=& \ 
\chi_{a \uparrow, b \uparrow; c \uparrow, d \uparrow}^{l l^{\prime}}(q)
- 
\chi_{a \downarrow, b \downarrow; c \uparrow, d \uparrow}^{l l^{\prime}}(q).\label{eq:chiS-def}
\end{align}  
These susceptibilities characterize the fluctuations in a PM state 
since the spin-orbit interaction is neglected. 
For example, 
the fluctuation for spin degrees of freedom (i.e., spin fluctuation) 
is characterized by 
\begin{align}
\chi^{\textrm{S}}(q)
=&\  
\dfrac{1}{2} 
\textstyle\sum\limits_{l,l^{\prime}=A,B}
\textstyle\sum\limits_{a,b=1}^{3}
\chi_{aabb}^{l l^{\prime} (\textrm{S})}(q),\label{eq:chiS-neutron}
\end{align}
where $1/2$ is the normalization constant for the summation 
with respect to the sublattice indices. 

$\chi_{abcd}^{l l^{\prime} (\textrm{C})}(q)$ and $\chi_{abcd}^{l l^{\prime} (\textrm{S})}(q)$ 
in Eqs. (\ref{eq:chiC-def}) and (\ref{eq:chiS-def}) 
are determined within the RPA 
by using the following equations:~\cite{RPA,RPA-Kubo,Kariyado-Mthesis} 
\begin{align}
\chi_{abcd}^{l l^{\prime} (\textrm{C})}(q)=&\ 
\chi_{abcd}^{l l^{\prime} (0)}(q)
+\textstyle\sum\limits_{\{a^{\prime}\}} 
\textstyle\sum\limits_{l^{\prime\prime}} 
\chi_{aba^{\prime}b^{\prime}}^{l l^{\prime\prime} (0)}(q) 
\Gamma_{a^{\prime}b^{\prime}c^{\prime}d^{\prime}}^{\textrm{C}}
\chi_{c^{\prime}d^{\prime}cd}^{l^{\prime\prime} l^{\prime} (\textrm{C})}(q),\label{eq:chiC-RPA}\\
\chi_{abcd}^{l l^{\prime} (\textrm{S})}(q)=&\ 
\chi_{abcd}^{l l^{\prime} (0)}(q)
+\textstyle\sum\limits_{\{a^{\prime}\}} 
\textstyle\sum\limits_{l^{\prime\prime}} 
\chi_{aba^{\prime}b^{\prime}}^{l l^{\prime\prime} (0)}(q) 
\Gamma_{a^{\prime}b^{\prime}c^{\prime}d^{\prime}}^{\textrm{S}}
\chi_{c^{\prime}d^{\prime}cd}^{l^{\prime\prime} l^{\prime} (\textrm{S})}(q),\label{eq:chiS-RPA}
\end{align}
where $\sum_{\{a^{\prime}\}}\equiv \sum_{a^{\prime},b^{\prime},c^{\prime},d^{\prime}}$ and 
the bare vertex interactions are 
\begin{align}
&\Gamma_{abcd}^{\textrm{C}}
= 
\begin{cases} 
\ -U \ \ \ \ \ \ \ \ \ \ \ \ \ \textrm{for} \ a=b=c=d\\
\ -2U^{\prime}+J_{\textrm{H}} \  \ \ \textrm{for} \ a=b\neq c=d\\
\ U^{\prime}-2J_{\textrm{H}} \ \  \ \ \ \textrm{for} \ a=d\neq b=c\\
\ -J^{\prime} \ \ \ \ \ \ \ \ \  \ \ \ \textrm{for} \ a=c\neq b=d\\
\ \ 0 \ \ \ \ \ \ \ \ \ \ \ \ \ \ \ \textrm{otherwise} \\
\end{cases},\label{eq:GammaC}\\
\hspace{-15pt}\textrm{and} &\notag\\
& \Gamma_{abcd}^{\textrm{S}}
= 
\begin{cases} 
\ U \ \ \ \ \ \ \ \textrm{for} \ a=b=c=d\\
\ J_{\textrm{H}} \ \ \ \ \ \ \textrm{for} \ a=b\neq c=d\\
\ U^{\prime} \ \ \ \ \ \ \textrm{for} \ a=d\neq b=c\\
\ J^{\prime} \ \ \ \ \ \ \ \textrm{for} \ a=c\neq b=d\\
\ \ 0 \ \ \ \ \ \ \  \textrm{otherwise} \\
\end{cases},\label{eq:GammaS}
\end{align}
respectively. 
These bare vertex interactions are independent of the sublattice indices 
since the interacting Hamiltonian (\ref{eq:Hint}) is diagonal 
in terms of these indices.

Since a phase transition is characterized 
by a divergence of static susceptibility, 
we calculate the eigenvalues of the static susceptibility 
by solving 
\begin{align}
\textstyle\sum\limits_{c,d}
\textstyle\sum\limits_{l^{\prime}}
\chi_{abcd}^{l l^{\prime} (\textrm{C})}(\boldq,0)
u_{cdl^{\prime};\nu}^{(\textrm{C})}(\boldq)
=&\
\lambda^{\textrm{C}}_{\nu}(\boldq)
u_{abl;\nu}^{(\textrm{C})}(\boldq),\\
\textstyle\sum\limits_{c,d}
\textstyle\sum\limits_{l^{\prime}}
\chi_{abcd}^{l l^{\prime} (\textrm{S})}(\boldq,0)
u_{cdl^{\prime};\nu}^{(\textrm{S})}(\boldq)
=&\
\lambda^{\textrm{S}}_{\nu}(\boldq)
u_{abl;\nu}^{(\textrm{S})}(\boldq).
\end{align}
Here, $\lambda^{\textrm{C}/\textrm{S}}_{\nu}(\boldq)$ 
and $u_{abl;\nu}^{(\textrm{C}/\textrm{S})}(\boldq)$ 
are the eigenvalue of static susceptibility for a charge/spin sector 
and the corresponding eigenvector. 
We analyze the instability from the maximum eigenvalues:~\cite{Tsunetsugu-RPA}
\begin{align}
\lambda^{\textrm{C}}_{\textrm{max}}(\boldq)
=&\max_{\nu}\biggl[
\textstyle\sum\limits_{\{a\}}
\textstyle\sum\limits_{l,l^{\prime}}
u_{\nu;abl}^{(\textrm{C}) \dagger}(\boldq)
\chi_{abcd}^{l l^{\prime} (\textrm{C})}(\boldq,0)
u_{cdl^{\prime};\nu}^{(\textrm{C})}(\boldq)\biggr],\label{eq:max-ev-chiC}\\
\lambda^{\textrm{S}}_{\textrm{max}}(\boldq)
=&\max_{\nu}\biggl[
\textstyle\sum\limits_{\{a\}}
\textstyle\sum\limits_{l,l^{\prime}}
u_{\nu;abl}^{(\textrm{S}) \dagger}(\boldq)
\chi_{abcd}^{l l^{\prime} (\textrm{S})}(\boldq,0)
u_{cdl^{\prime};\nu}^{(\textrm{S})}(\boldq)\biggr].\label{eq:max-ev-chiS}
\end{align} 

Finally, 
to address the role of each Ru $t_{2g}$ orbital 
in the fluctuations, 
we introduce the susceptibilities averaged 
with respect to the sublattice indices:
\begin{align}
\chi_{abcd}^{(0)}(q)
=&\ \dfrac{1}{2} 
\textstyle\sum\limits_{l,l^{\prime}=A,B}
\chi_{abcd}^{l l^{\prime} (0)}(q),\label{eq:av-chi0}\\
\chi_{abcd}^{(\textrm{C})}(q)
=&\ \dfrac{1}{2} 
\textstyle\sum\limits_{l,l^{\prime}=A,B}
\chi_{abcd}^{l l^{\prime} (\textrm{C})}(q),\label{eq:av-chiC}\\
\chi_{abcd}^{(\textrm{S})}(q)
=&\ \dfrac{1}{2} 
\textstyle\sum\limits_{l,l^{\prime}=A,B}
\chi_{abcd}^{l l^{\prime} (\textrm{S})}(q).\label{eq:av-chiS}
\end{align}

\section{Results}
In the following calculations, 
we set $T=0.02$, $J^{\prime}=J_{\textrm{H}}$, and $U^{\prime}=U-2J_{\textrm{H}}$, 
and use the values of $U$ and $J_{\textrm{H}}$ as parameters. 
To calculate the static susceptibilities, 
we divide the BZ into 128$\times$128 meshes and 
take $1024$ fermionic Matsubara frequencies 
for using the fast Fourier transformation. 

\subsection{Effects of the rotation of RuO$_{6}$ octahedra}
\begin{figure*}[tb]
\vspace{-4pt}
\includegraphics[width=172mm]{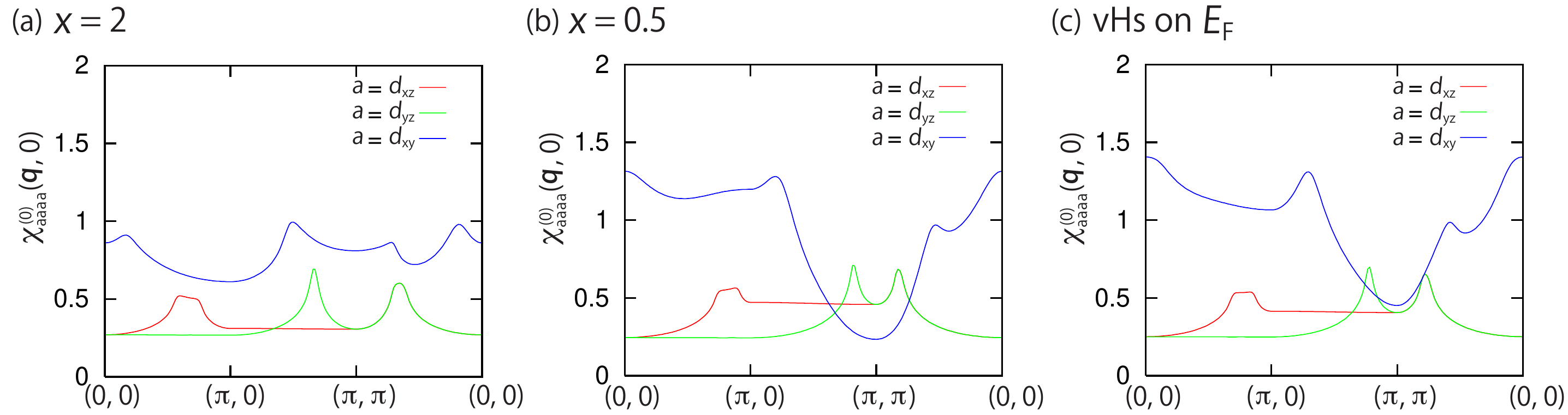}
\vspace{-5pt}
\caption{(Color online) 
Momentum dependencies of $\chi_{aaaa}^{(0)}(\boldq,0)$ 
for the models of (a) $x=2$ and (b) $0.5$, and (c) the special model. }
\label{fig:chi0}
\end{figure*}
\begin{figure*}[tb]
\vspace{10pt}
\includegraphics[width=172mm]{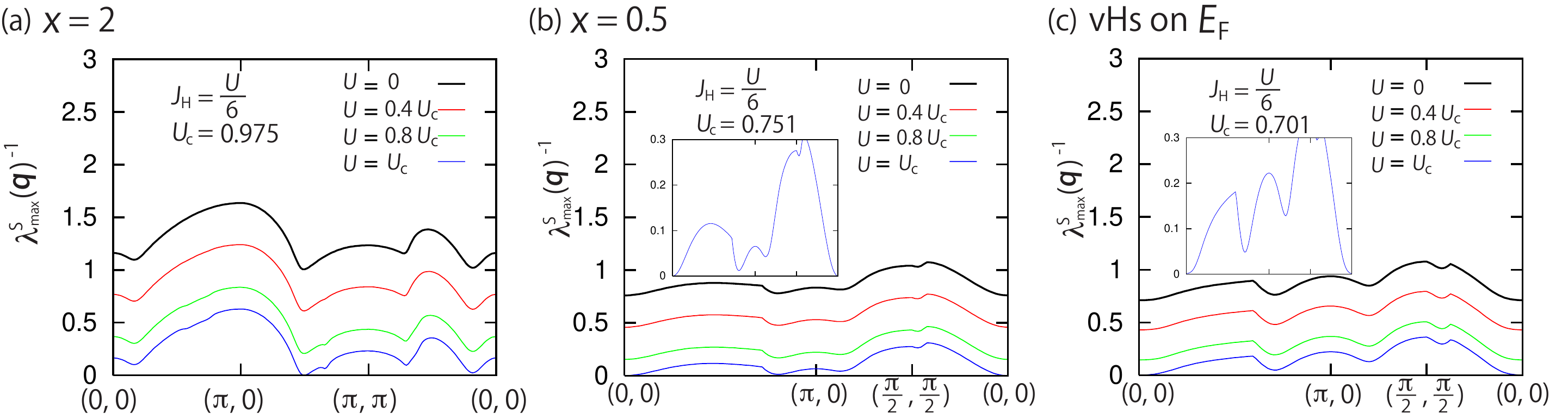}
\vspace{-5pt}
\caption{(Color online) 
Momentum dependencies of $\lambda^{\textrm{S}}_{\textrm{max}}(\boldq)^{-1}$ 
in the RPA for the models of (a) $x=2$ and (b) $0.5$, 
and (c) the special model. 
The insets in (b) and (c) show 
the momentum dependencies of $\lambda^{\textrm{S}}_{\textrm{max}}(\boldq)^{-1}$ 
at $U=U_{\textrm{c}}$ in an expanded scale 
for the model of $x=0.5$ and the special model, respectively.}
\label{fig:evS-JH0167}
\end{figure*}
\begin{figure*}[tb]
\vspace{-4pt}
\includegraphics[width=172mm]{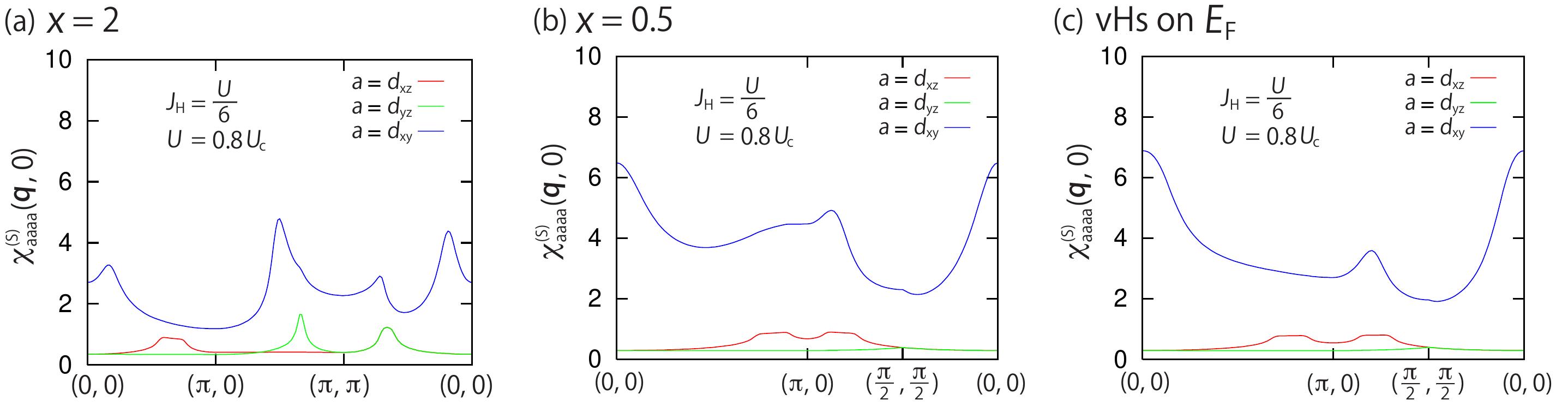}
\vspace{-5pt}
\caption{(Color online) 
Momentum dependencies of $\chi_{aaaa}^{(\textrm{S})}(\boldq,0)$ 
in the RPA for the models of (a) $x=2$ and (b) $0.5$, 
and (c) the special model. }
\label{fig:chiS-JH0167}
\end{figure*}
\begin{figure}[tb]
\includegraphics[width=86mm]{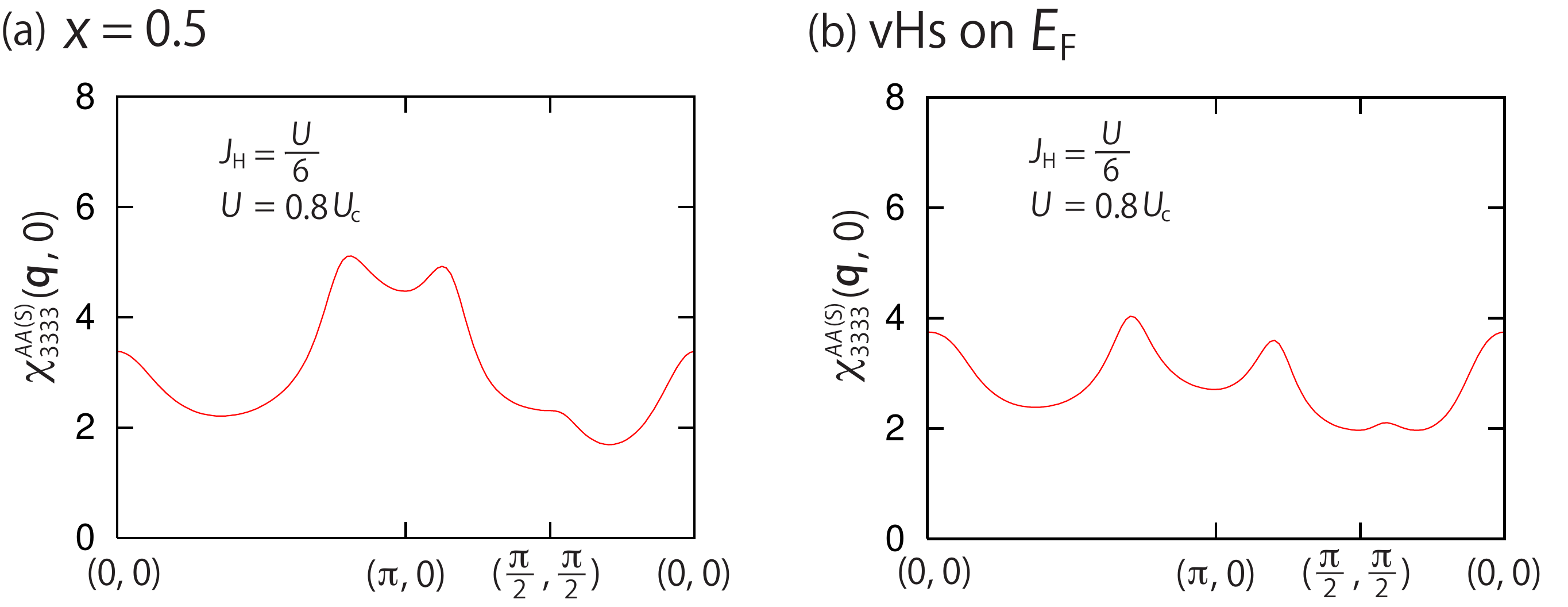}
\vspace{-22pt}
\caption{(Color online) 
Momentum dependencies of $\chi_{aaaa}^{A A(\textrm{S})}(\boldq,0)$ 
for the $d_{xy}$ orbital in the RPA 
for (a) the model of $0.5$ and (b) the special model. }
\label{fig:chiS-sublattice-JH0167}
\end{figure}
\begin{figure*}[tb]
\vspace{10pt}
\includegraphics[width=172mm]{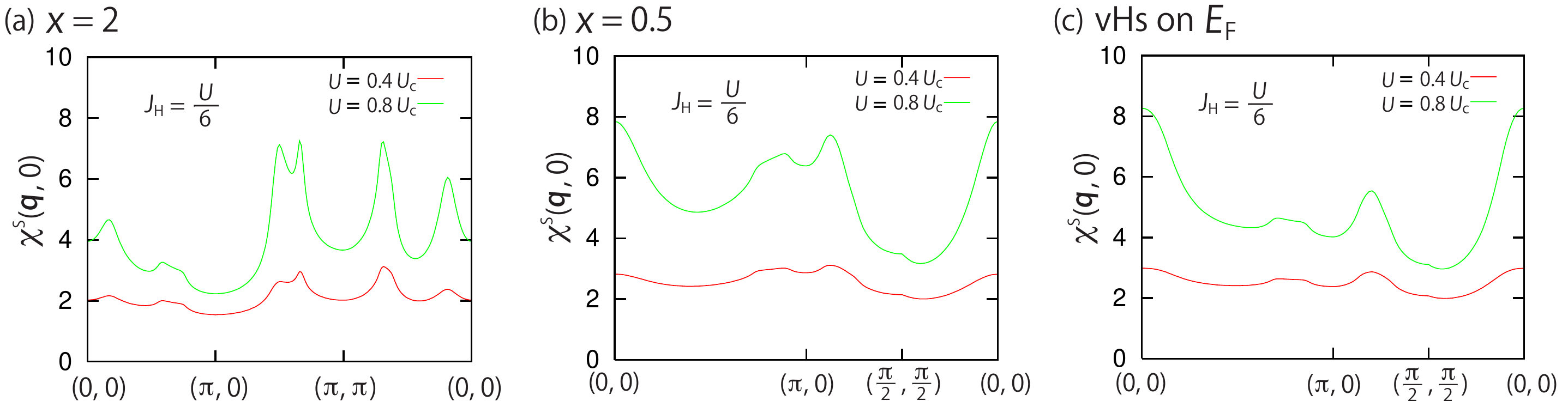}
\vspace{-5pt}
\caption{(Color online) 
Momentum dependencies of $\chi^{\textrm{S}}(\boldq,0)$ 
in the RPA for the models of (a) $x=2$ and (b) $0.5$, 
and (c) the special model. }
\label{fig:chiSav-JH0167}
\end{figure*}

To analyze the effects of the rotation of RuO$_{6}$ octahedra 
on the static susceptibilities, 
we compare the results for the model of $x=2$ (i.e., $\phi=0^{\circ}$) 
and $0.5$ (i.e., $\phi=15^{\circ}$). 

First, the momentum dependencies of 
the noninteracting susceptibility for these models 
are shown in Figs. \ref{fig:chi0} (a) and (b). 
[Note that the momentum lines in both figures are along 
the arrows shown in Fig. \ref{fig:FS} (b) 
to compare the cases of $x=2$ and $0.5$.] 
We find that 
the noninteracting susceptibility for the $d_{xy}$ orbital around $\boldq=(0,0)$ 
is larger for the model of $x=0.5$ than for the model of $x=2$. 
This difference is due to the increase of the DOS 
for the $d_{xy}$ orbital near the Fermi level 
in the presence of the rotation of RuO$_{6}$ octahedra.~\cite{NA-GA} 
We also find that 
the incommensurate (IC) peaks for the $d_{xz/yz}$ orbital 
around $\boldq=(\pi,\pi)$ 
approach $\boldq=(\pi,\pi)$ 
as $x$ is changed from $2$ to $0.5$; 
simultaneously, the IC peak for the $d_{xy}$ orbital around $\boldq=(\pi,0)$ 
approach $\boldq=(\pi,0)$. 
These shifts are understood as follows: 
The rotation-induced hybridization of the $d_{xy}$ orbital to 
the $d_{x^{2}\textrm{-}y^{2}}$ orbital leads to the downward shift of 
the energy level for the $d_{xy}$ orbital.~\cite{NA-GA} 
Correspondingly, the energy level for the $d_{xz/yz}$ orbital 
goes up compared with that for the $d_{xy}$ orbital. 
The upward shift of the band for the $d_{xz/yz}$ orbital 
around $\boldq=(\pi,\pi)$ crossing the Fermi level leads to 
an increase of the nesting vector towards $\boldq=(\pi,\pi)$. 

Next, we show the momentum dependencies of the maximum eigenvalue 
of static susceptibility in the RPA at $J_{\textrm{H}}=U/6$ 
for the models of $x=2$ and $0.5$. 
Figure \ref{fig:evS-JH0167} (a) shows the momentum dependencies 
of $\lambda^{\textrm{S}}_{\textrm{max}}(\boldq)^{-1}$ for the model of $x=2$. 
We find that 
$\lambda^{\textrm{S}}_{\textrm{max}}(\boldq)^{-1}$ at $\boldq\sim (\pi,0.5\pi)$ 
touches zero at $U=U_{\textrm{c}}=0.975$, 
and that 
there are two main peaks at $\boldq\sim (\pi,0.5\pi)$ 
and $(0.235\pi,0.235\pi)$ 
and two secondary peaks at $\boldq\sim(\pi,0.656\pi)$ 
and $(0.688\pi,0.688\pi)$. 
Note that 
the enhancement of these main peaks is inconsistent with 
the experimental result of 
an inelastic neutron measurement,~\cite{Neutron-Sr214} 
in which 
the enhancement of the IC antiferromagnetic (AF) spin fluctuation 
with $\boldq\sim(0.6\pi,0.6\pi)$ is observed. 
However, as we will show below, our model can explain this experimental result. 
We also find that 
the increase of $U$ leads to the suppression 
of the values of $\lambda^{\textrm{C}}_{\textrm{max}}(\boldq)^{-1}$ (not shown here). 

It should be noted that 
the small value of $U_{\textrm{c}}$ 
is due to the overestimation of the fluctuations in the RPA. 
This small value of $U_{\textrm{c}}$ will be unrealistic 
since the value of $U_{\textrm{c}}$ for transition metal oxides 
will be of the order of $2-3$ eV. 
The similar problem remains for the model of $x=0.5$ and the special model, 
as we will show below. 
In Sec. IV, we will address this problem and 
effects of the terms neglected in the RPA. 

The momentum dependencies of $\lambda^{\textrm{S}}_{\textrm{max}}(\boldq)^{-1}$ 
for the model of $x=0.5$ are shown in Fig. \ref{fig:evS-JH0167} (b); 
the inset represents $\lambda^{\textrm{S}}_{\textrm{max}}(\boldq)^{-1}$ 
at $U=U_{\textrm{c}}=0.751$ in an expanded scale. 
[Hereafter, the momentum lines for $\phi\neq 0^{\circ}$ 
are along the arrows shown in Fig. \ref{fig:FS} (a).] 
We find that 
$\lambda^{\textrm{S}}_{\textrm{max}}(\boldq)^{-1}$ 
at $\boldq=(0,0)$ touches zero at $U=U_{\textrm{c}}=0.751$, 
which is smaller than for the model of $x=2$, 
and that 
there are two main peaks at $\boldq=(0,0)$ and $\boldq\sim (0.797\pi)$ 
and a secondary peak at $\boldq\sim(\pi,0.125\pi)$; 
all these peaks are less sharp than the peaks for the model of $x=2$. 
As we will discuss in Sec. IV B, 
the enhancement of the main peak at $\boldq=(0,0)$ 
is consistent with several experiments.~\cite{Ishida-NMR-x05,Neutron-gamma,
Neutron-New} 
We also find that 
the values of $\lambda^{\textrm{C}}_{\textrm{max}}(\boldq)^{-1}$ (not shown here) 
suppress as $U$ increases. 
These results are qualitatively the same as 
those obtained in the previous study,~\cite{NA-ICM} 
where the NNN hopping integral between the $d_{xz}$ and $d_{yz}$ orbitals 
has been neglected. 

Moreover, Figs. \ref{fig:chiS-JH0167} (a) and (b) 
show the momentum dependencies of 
$\chi_{aaaa}^{(\textrm{S})}(\boldq,0)$ in the RPA at $U=0.8U_{\textrm{c}}$ 
for the models of $x=2$ and $0.5$. 
For the model of $x=2$, 
we see from Figs. \ref{fig:evS-JH0167} (a) and \ref{fig:chiS-JH0167} (a) 
that 
the two main peaks in $\lambda^{\textrm{S}}_{\textrm{max}}(\boldq)^{-1}$ arise from 
the corresponding fluctuations for the $d_{xy}$ orbital, 
and that the secondary peaks in $\lambda^{\textrm{S}}_{\textrm{max}}(\boldq)^{-1}$ 
at $\boldq\sim(\pi,0.656\pi)$ and $(0.688\pi,0.688\pi)$ arise from 
the corresponding fluctuation for the $d_{yz}$ orbital and 
the combined fluctuation of the $d_{xz/yz}$ and $d_{xy}$ orbitals. 
These results suggest that 
the fluctuations for a spin sector 
not only for the $d_{xy}$ orbital but also for the $d_{xz/yz}$ orbital 
play an important role for $x=2$. 

For the model of $x=0.5$, 
we see from Figs. \ref{fig:evS-JH0167} (b), \ref{fig:chiS-JH0167} (b), 
and \ref{fig:chiS-sublattice-JH0167} (a) 
that 
the two main peaks and the secondary peak 
in $\lambda^{\textrm{S}}_{\textrm{max}}(\boldq)^{-1}$ 
arise from the corresponding fluctuations for the $d_{xy}$ orbital, 
and that the contribution from the $d_{xz/yz}$ orbital 
is much smaller than for the $d_{xy}$ orbital. 
Note that the value of $\chi_{aaaa}^{AA(\textrm{S})}(\boldq,0)$ 
for the $d_{xy}$ orbital is about $10$ times larger than 
for the $d_{xz/yz}$ orbital. 
These results suggest that 
the rotation of RuO$_{6}$ octahedra leads to 
the enhancement of several modes of spin fluctuation 
for the $d_{xy}$ orbital around $\boldq=(0,0)$ and $\boldq\sim (\pi,0)$, 
and that the fluctuation for a spin sector for the $d_{xy}$ orbital 
plays a more important role for $x=0.5$ than for the $d_{xz/yz}$ orbital. 

Finally, 
Figs. \ref{fig:chiSav-JH0167} (a) and (b) show 
the momentum dependencies of $\chi^{\textrm{S}}(\boldq,0)$ 
in the RPA at $J_{\textrm{H}}=U/6$ 
for $U=0.8U_{\textrm{c}}$ and $0.4U_{\textrm{c}}$ for the models of $x=2$ and $0.5$. 
For the model of $x=2$, 
we find from Fig. \ref{fig:chiSav-JH0167} (a) 
that 
$\chi^{\textrm{S}}(\boldq,0)$ at $\boldq\sim(\pi,0.5\pi)$, 
$(\pi,0.656\pi)$, and $(0.688\pi,0.688\pi)$ are dominant at $U=0.8U_{\textrm{c}}$, 
while 
$\chi^{\textrm{S}}(\boldq,0)$ at $\boldq\sim (0.688\pi,0.688\pi)$ 
becomes largest at $U=0.4U_{\textrm{c}}$; 
the main peak in $\lambda^{\textrm{S}}_{\textrm{max}}(\boldq)^{-1}$ 
at $\boldq\sim (0.235\pi,0.235\pi)$ is less important 
at $U=0.8U_{\textrm{c}}$ and $0.4U_{\textrm{c}}$. 
These results suggest that 
the experimentally observed enhancement~\cite{Neutron-Sr214} 
of the IC AF spin fluctuation at $\boldq\sim(0.6\pi,0.6\pi)$ 
can be understood if Sr$_{2}$RuO$_{4}$ is not located in the vicinity 
of the magnetic order. 
As we will show in Sec. III C, 
the Hund's rule coupling also plays an important role 
in enhancing this IC AF spin fluctuation. 

For the model of $x=0.5$, 
we find from Fig. \ref{fig:chiSav-JH0167} (b) 
that $\chi^{\textrm{S}}(\boldq,0)$ at $\boldq=(0,0)$ 
and $\boldq\sim(\pi,0.125\pi)$ 
are dominant at $U=0.8U_{\textrm{c}}$, 
while $\chi^{\textrm{S}}(\boldq,0)$ at $\boldq=(0,0)$, 
$\boldq\sim(0.797\pi,0)$, and $(\pi,0.125\pi)$ 
are nearly the same at $U=0.4U_{\textrm{c}}$. 
Namely, almost all contributions arise from 
the corresponding fluctuations for the $d_{xy}$ orbital. 
We also find that 
the values of $\chi^{\textrm{S}}(\boldq,0)$ 
along $(0,0)\rightarrow (\pi,0)$ 
are larger than along $(0,0)\rightarrow (\pi/2,\pi/2)$. 
This result indicates that 
the spin fluctuations along $(0,0)\rightarrow (\pi,0)$ are strongly enhanced 
for $x=0.5$. 
This result is consistent with several inelastic neutron 
measurements,~\cite{Neutron-gamma,Neutron-New} 
as we will address in Sec. IV B. 
These results suggest that 
the spin fluctuations for the $d_{xy}$ orbital 
along $(0,0)\rightarrow (\pi,0)$ 
play a very important role for $x=0.5$. 

\subsection{Effects of the vHs for the $d_{xy}$ orbital}
To analyze the effects of the vHs for the $d_{xy}$ orbital 
on the static susceptibilities, 
we compare the results for the model of $x=0.5$ 
with those for the special model, 
where the vHs is located on the Fermi level; 
the difference in these models is only the value of $\Delta_{t_{2g}}$. 

First, the momentum dependence of 
the noninteracting susceptibility for the special model 
is shown in Fig. \ref{fig:chi0} (c). 
Comparing this figure with Fig. \ref{fig:chi0} (b), 
we see that 
the noninteracting susceptibility for the $d_{xy}$ orbital around $\boldq=(0,0)$ 
is larger for the special model than for the model of $x=0.5$. 
This difference is due to 
the increase of the DOS for the $d_{xy}$ orbital near the Fermi level 
in the special model. 
We also see that 
the shifts of the IC peak for the $d_{xz/yz}$ and $d_{xy}$ orbitals, 
which are observed in the model of $x=0.5$, 
are smaller in the special model. 
This difference arises from the smaller value of $\Delta_{t_{2g}}$ 
for the special model.  

Next, Fig. \ref{fig:evS-JH0167} (c) shows the momentum dependencies of 
$\lambda^{\textrm{S}}_{\textrm{max}}(\boldq)^{-1}$ in the RPA at $J_{\textrm{H}}=U/6$ 
for the special model; 
the inset represents $\lambda^{\textrm{S}}_{\textrm{max}}(\boldq)^{-1}$ 
at $U=U_{\textrm{c}}=0.701$ in an expanded scale. 
Comparing Fig. \ref{fig:evS-JH0167} (c) with 
Fig. \ref{fig:evS-JH0167} (b), 
we find that 
the value of $U_{\textrm{c}}$, 
where $\lambda^{\textrm{S}}_{\textrm{max}}(\boldq)^{-1}$ 
at $\boldq=(0,0)$ touches zero, 
is smaller for the special model than for the model of $x=0.5$, 
and that 
there is a main peak at $\boldq=(0,0)$ and there is a secondary peak 
at $\boldq\sim(0.703\pi,0)$; 
the latter peak corresponds to the peak at $\boldq\sim(0.797\pi,0)$ 
for the model of $x=0.5$. 
In addition, 
we see that 
all the peaks in $\lambda^{\textrm{S}}_{\textrm{max}}(\boldq)^{-1}$ 
for the special model are slightly sharper than for the model of $x=0.5$, 
and that 
the competition between the modes of 
spin fluctuation around $\boldq=(0,0)$ and $\boldq\sim (\pi,0)$ 
is weaker for the special model. 
The latter is due to the larger increase of the noninteracting 
susceptibilities for the $d_{xy}$ orbital 
around $\boldq=(0,0)$ compared with those around $\boldq\sim (\pi,0)$ 
for the special model than for the model of $x=0.5$. 
These results indicate that 
the location of the vHs for the $d_{xy}$ orbital, 
which is controlled by substitution of Ca for Sr, 
is the parameter to control this competition. 
Note that 
the momentum dependencies of $\lambda^{\textrm{C}}_{\textrm{max}}(\boldq)^{-1}$ 
and $\chi_{aaaa}^{\textrm{C}}(\boldq,0)$ (not shown) are qualitatively the same to 
those obtained for the model of $x=0.5$. 

Finally, Figs. \ref{fig:chiS-JH0167} (c), 
\ref{fig:chiS-sublattice-JH0167} (b), 
and \ref{fig:chiSav-JH0167} (c) show 
the momentum dependencies of 
$\chi_{aaaa}^{(\textrm{S})}(\boldq,0)$, $\chi_{3333}^{AA(\textrm{S})}(\boldq,0)$, 
and $\chi^{\textrm{S}}(\boldq,0)$ in the RPA for the special model.  
Comparing these figures with Figs. \ref{fig:chiS-JH0167} (b), 
\ref{fig:chiS-sublattice-JH0167} (a), and \ref{fig:chiSav-JH0167} (b), 
we find the qualitatively same results to those for the model of $x=0.5$. 
Namely, 
the enhanced modes in $\lambda^{\textrm{S}}_{\textrm{max}}(\boldq)^{-1}$ 
arise from the corresponding fluctuations 
for the $d_{xy}$ orbital, 
and the dominant contributions to $\chi_{aaaa}^{(\textrm{S})}(\boldq,0)$ 
arise from those for the $d_{xy}$ orbital around $\boldq=(0,0)$ 
and $\boldq\sim (\pi,0)$. 
These results suggest that 
the spin fluctuation for the $d_{xy}$ orbital is more important 
in the electronic states around $x=0.5$ than for the $d_{xz/yz}$ orbital. 

\subsection{Effects of the Hund's rule coupling}
\begin{figure}[tb]
\includegraphics[width=88mm]{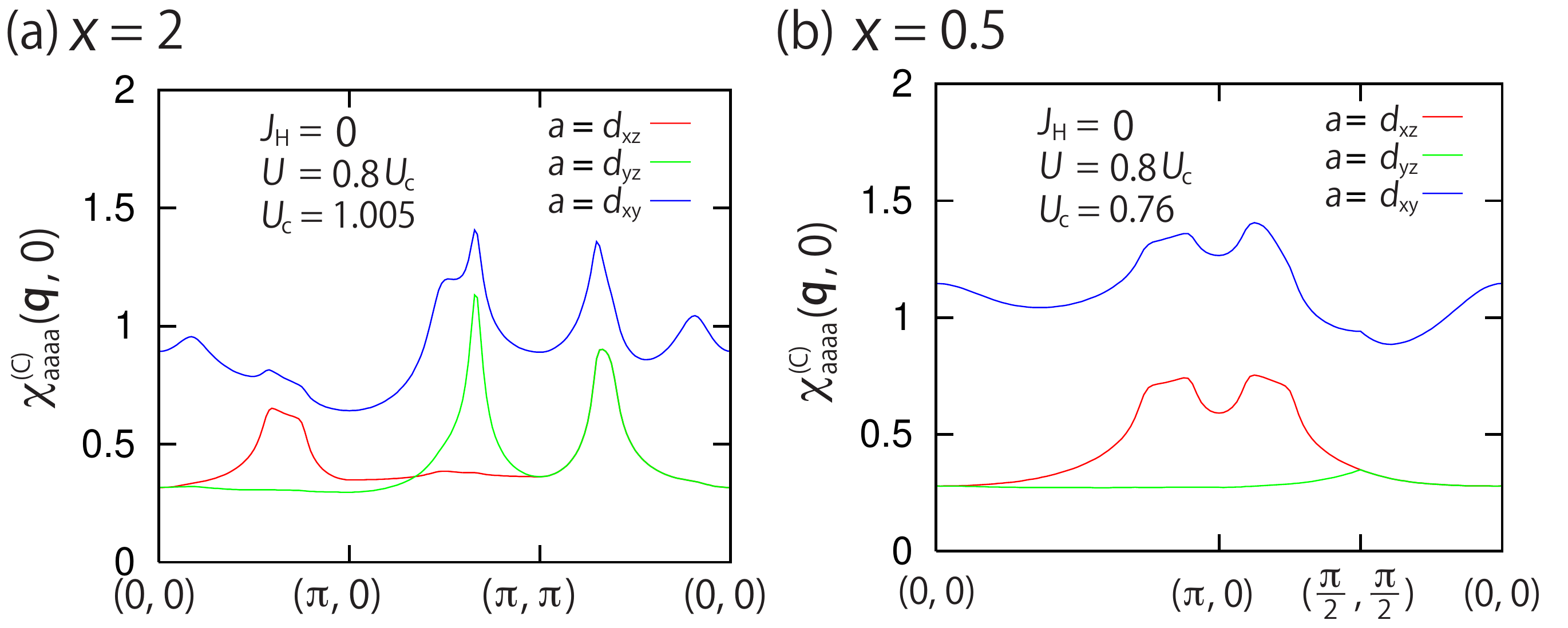}
\vspace{-16pt}
\caption{(Color online) 
Momentum dependencies of $\chi_{aaaa}^{(\textrm{C})}(\boldq,0)$ 
in the RPA for the models of (a) $x=2$ and (b) $0.5$. }
\label{fig:RPA-chiC-JH0}
\end{figure}
\begin{figure*}[tb]
\vspace{-4pt}
\includegraphics[width=172mm]{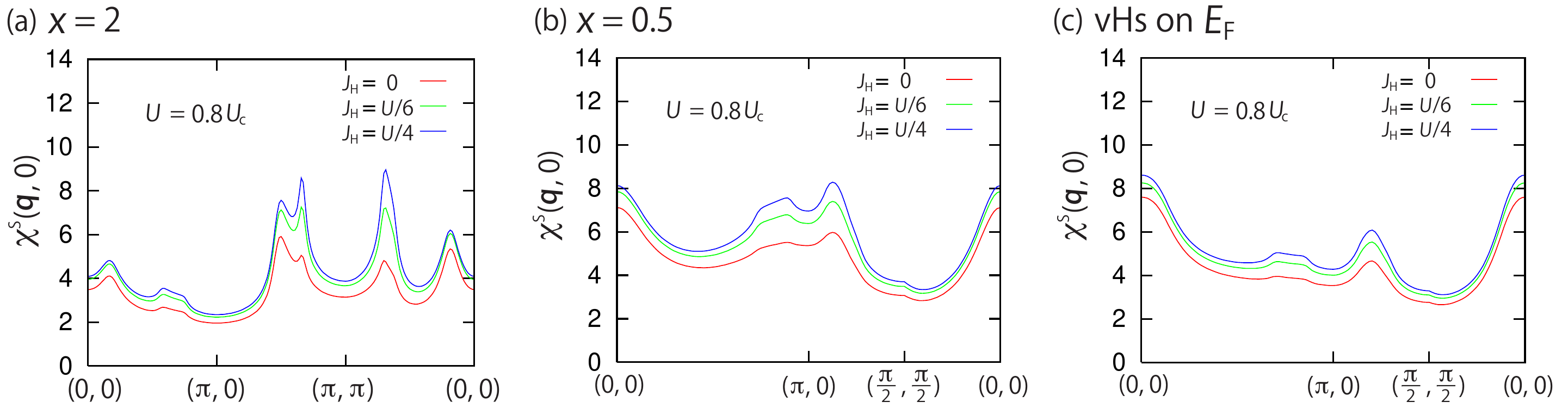}
\vspace{-5pt}
\caption{(Color online) 
Momentum dependencies of $\chi^{\textrm{S}}(\boldq,0)$ in the RPA 
for the models of (a) $x=2$ and (b) $x=0.5$, and (c) the special model. }
\label{fig:chiS-av}
\end{figure*}

To analyze the effects of the Hund's rule coupling on 
the static susceptibilities, 
we compare the results in the RPA at $J_{\textrm{H}}=U/6$ 
with those at $J_{\textrm{H}}=0$ and $U/4$. 

First, we focus on the $J_{\textrm{H}}$ dependence of 
the static susceptibilities for the model of $x=2$. 
We find that 
the dominant wave vectors of these susceptibilities 
at $J_{\textrm{H}}=0$ and $U/4$ (not shown) are same to 
those at $J_{\textrm{H}}=U/6$, 
and that 
the increase of $J_{\textrm{H}}/U$ leads to the decrease of $U_{\textrm{c}}$. 
The latter is a typical behavior for a multiorbital system 
since the increase of $J_{\textrm{H}}/U$ leads to 
an enhancement of spin fluctuations. 
We also find from Fig. \ref{fig:RPA-chiC-JH0} (a) that 
the static susceptibility for a charge sector 
is enhanced only at $J_{\textrm{H}}=0$ as $U$ increases. 
The similar enhancement has been reported in a RPA analysis for $x=2$,~\cite{Takimoto-RPA} 
where the NNN hopping integral between the $d_{xz}$ and $d_{yz}$ orbitals 
has been neglected. 
Furthermore, 
we see from Fig. \ref{fig:chiS-av} (a) that 
$\chi^{\textrm{S}}(\boldq,0)$ around $\boldq\sim (0.688\pi,0.688\pi)$ 
and $(\pi,0.656\pi)$ 
are strongly enhanced as $J_{\textrm{H}}/U$ increases; 
these modes correspond to the secondary peaks 
in $\lambda^{\textrm{S}}_{\textrm{max}}(\boldq)^{-1}$ at $J_{\textrm{H}}=U/6$. 
The nature of these enhancements can be understood 
as follows:~\cite{Nomura-MFA} 
the Hund's rule coupling, 
which is one of the interorbital Coulomb interactions, 
more affects the orbitals having a finite hybridization. 

Next, we turn to the $J_{\textrm{H}}$ dependence of 
the static susceptibilities for the model of $x=0.5$. 
Similarly to the case of $x=2$, 
we find that the dominant wave vectors of the static susceptibilities at 
$J_{\textrm{H}}=0$ and $U/4$ (not shown) are same to those at $J_{\textrm{H}}=U/6$, 
that the increase of $J_{\textrm{H}}/U$ leads to the decrease of $U_{\textrm{c}}$, 
and that the increase of $U$ leads to the enhancement of 
the static susceptibility for a charge sector only at $J_{\textrm{H}}=0$ 
[see Fig. \ref{fig:RPA-chiC-JH0} (b)]. 
In contrast to the case of $x=2$, 
we find from Fig. \ref{fig:chiS-av} (b) that 
the increase of $J_{\textrm{H}}/U$ leads to 
a nearly uniform enhancement of $\chi^{\textrm{S}}(\boldq,0)$; 
more precisely,  
the enhancement of $\chi^{\textrm{S}}(\boldq,0)$ around 
$\boldq\sim (\pi,0)$ is slightly larger than 
around $\boldq=(0,0)$. 
This difference is due to 
the much larger contribution from the $d_{xy}$ orbital 
for the model of $x=0.5$ than from the $d_{xz/yz}$ orbital. 
These results indicate that 
the static susceptibilities for a spin sector for the $d_{xy}$ orbital 
are very important for $x=0.5$ 
regardless of the value of $J_{\textrm{H}}/U$. 

Before showing the results for the special model, 
let us remark on the effect of 
the rotation-induced hybridization of the $d_{xy}$ orbital 
to the $d_{x^{2}{\textrm{-}}y^{2}}$ orbital 
on the $J_{\textrm{H}}$ dependence of $\chi^{\textrm{S}}(\boldq,0)$ 
for the model of $x=0.5$. 
In this study, 
the effect of this hybridization is approximately taken account 
as the difference of the CEF energies between the $d_{xz/yz}$ 
and $d_{xy}$ orbitals [see Eqs. (\ref{eq:dis11AA-phi}) 
and (\ref{eq:dis33AA-phi})]. 
In principle, 
if this hybridization is fully taken into account, 
the $J_{\textrm{H}}$ dependence of $\chi^{\textrm{S}}(\boldq,0)$ will be modified. 
However, 
according to the density-functional calculation for $x=0.5$,~\cite{Oguchi-LS} 
the $d_{x^{2}{\textrm{-}}y^{2}}$ orbital has 
little weights to the bands near the Fermi level for $x=0.5$. 
This result indicates that 
the values of the static susceptibility for the $d_{x^{2}\textrm{-}y^{2}}$ orbital 
are much smaller compared with those for the $t_{2g}$ orbitals. 
Thus, the $J_{\textrm{H}}$ dependence of $\chi^{\textrm{S}}(\boldq,0)$ 
does not qualitatively change from the present result 
if this rotation-induced hybridization is fully taken into account.   

Finally, we focus on the $J_{\textrm{H}}$ dependence of 
the static susceptibilities for the special model. 
We find the similar $J_{\textrm{H}}$ dependencies 
of the value of $U_{\textrm{c}}$ and the static susceptibilities 
to those for the model of $x=0.5$ [e.g., see 
Fig. \ref{fig:chiS-av} (c)]. 
Therefore, we conclude that 
the fluctuations for a spin sector for the $d_{xy}$ orbital 
play a very important role in the electronic states around $x=0.5$. 

\section{Discussion}
We first address the mass enhancement due to the enhanced fluctuations 
for the models of $x=2$ and $0.5$. 
In this study, we find that 
the value of $U_{\textrm{c}}$ is smaller for the model of $x=0.5$ 
than for the model of $x=2$, 
and that 
the peaks in $\lambda^{\textrm{S}}_{\textrm{max}}(\boldq)^{-1}$ are less sharp 
for the model of $x=0.5$ [see Figs. \ref{fig:evS-JH0167} (a) and (b)]. 
In general, 
the effective mass at a certain value of $U$ becomes large 
if the value of $U_{\textrm{c}}$ is small.~\cite{NearlyFM1,NearlyFM2,NearlyAF,
Ueda-paramag} 
In addition, 
the effective mass becomes larger for a less sharp peak in the susceptibility, 
even if the value of the peak is same;~\cite{NearlyFM1,NearlyFM2,NearlyAF,
Ueda-paramag} 
this arises from the product of 
the susceptibility and 
the summation with respect to $\boldq$ in the free energy. 
Combining these with our results, 
we conclude that 
the mass enhancement for the model of $x=0.5$ 
will be much larger than for the model of $x=2$. 
Namely, we think that 
the present results can qualitatively explain the experimentally 
observed mass enhancement towards $x=0.5$ (Ref.\onlinecite{Nakatsuji-HF}); 
the actual calculation about the mass enhancement is a remaining future work. 
 
\subsection{Comparison with previous theoretical studies}
We first compare our results 
with the previous theoretical studies about the magnetic properties 
for Ca$_{2-x}$Sr$_{x}$RuO$_{4}$ with $x=2$. 
The mean-field calculation~\cite{Nomura-MFA} for $x=2$ 
has found that 
the dominant contribution to the static susceptibility arises from 
that for the $d_{xy}$ orbital. 
This calculation has also shown that 
the increase of $J_{\textrm{H}}/U$ leads to the enhancement of 
the IC AF spin fluctuation 
around $\boldq\sim (0.67\pi,0.67\pi)$, 
and that this IC AF spin fluctuation is strongest 
only for $J_{\textrm{H}}\geq U/2$ at $U=1.25t_{3}\sim 0.56$. 
Note that this mode of the enhanced IC AF spin fluctuation 
is different from $\boldq\sim(0.688\pi,0.688\pi)$, 
which is enhanced in our study, due to the difference in the values 
of the hopping integrals. 
These results suggest that 
a moderately large Hund's rule coupling plays an important role for $x=2$. 
This is consistent with 
the previous dynamical mean-field theory~\cite{Haule-DMFT} for $x=2$. 
Similarly to this mean-field study,~\cite{Nomura-MFA} 
we find that 
the increase of $J_{\textrm{H}}/U$ leads to the enhancement of 
the IC AF spin fluctuations around $\boldq\sim (0.688\pi,0.688\pi)$ 
and $(\pi,0.656\pi)$; 
the former IC AF spin fluctuation is strongest for $J_{\textrm{H}}\geq U/6$ 
at $U=0.8U_{\textrm{c}}\sim 0.78$. 
Since the enhancement of the IC AF spin fluctuation 
at $\boldq\sim (0.6\pi,0.6\pi,0)$ 
has been experimentally observed,~\cite{Neutron-Sr214} 
we think that a set of the parameters used in this study 
is more realistic than 
in this previous mean-field study (Ref. \onlinecite{Nomura-MFA}). 

Next, we address the physical meaning of the results obtained in the RPA, 
and compare our results with the studies 
in more elaborated treatments. 
The RPA is a mean-field-type approximation 
and neglects the mode-mode coupling for fluctuations. 
In general, 
the mode-mode coupling plays an important role 
in discussing the electronic states 
near a quantum critical point (QCP).~\cite{Moriya-SCR,Moriya-SCR-AF,
Moriya-review,Moriya-CW} 
For example, 
it weakens the instability obtained in the RPA. 
Actually, the previous study~\cite{Yanase} for $x=2$ 
in fluctuation exchange (FLEX) approximation 
has observed much larger $U_{\textrm{c}}$ than in the RPA; 
the FLEX approximation is partially taken account of the mode-mode coupling. 
We thus expect that the values of $U_{\textrm{c}}$ 
will be more realistic than those obtained in this paper. 

In addition to this, 
the mode-mode coupling can change the primary instability 
from that obtained in the RPA. 
Thus, the results about the primary instability 
may change when the mode-mode coupling is taken into account. 
Actually, the previous study~\cite{Yanase} for $x=2$ in the FLEX approximation 
found that the IC AF spin fluctuation 
with $\boldq\sim (0.67\pi,0.67\pi)$ is very strong, 
while the spin fluctuations 
with $\boldq\sim(\pi,0.5\pi)$ and $(0.235\pi,0.235\pi)$ 
are strongly enhanced in the RPA; 
the reason why this mode of the enhanced IC AF spin fluctuation 
in the FLEX is different from that in our study 
is the same as that for a case of the difference between 
the previous mean-field study~\cite{Nomura-MFA} and ours. 
However, we believe that 
the competition between the modes of spin fluctuation 
around $\boldq=(0,0)$ and $\boldq\sim (\pi,0)$ 
for the model of $x=0.5$, which is obtained in this study, 
will not change 
even in a more elaborated treatment 
since the momentum dependence of $\lambda^{\textrm{S}}_{\textrm{max}}(\boldq)^{-1}$ 
is very flat; in such a case, it will be difficult that 
the mode-mode coupling favors a specific mode. 
Actually, our preliminary calculation~\cite{NA-FLEX} 
in the FLEX approximation observes the similar competition 
of spin fluctuations around these modes for the model of $x=0.5$. 

There are several density-functional calculations to study 
the ground states for $0.5\leq x \leq 2$. 
Among them, a density-functional calculation 
within local-density approximation (LDA) has found that 
the rotation of RuO$_{6}$ octahedra enhances 
the ferromagnetic (FM) instability.~\cite{Terakura}  
Another density-functional calculation~\cite{hyb-t2g-eg} for $x=0.5$ 
within the LDA has predicted the nesting instability 
with $\boldq\sim(0.29\pi,0.29\pi,0)$. 
The authors have proposed that this nesting instability will lead to 
the mass enhancement around $x=0.5$; 
this nesting vector is related to a new FS induced by 
the rotation-induced hybridization of the $d_{xy}$ orbital 
to the $d_{x^{2}\textrm{-}y^{2}}$ orbital. 
In our study, 
this new FS does not appear 
since we have not treated directly 
the rotation-induced hybridization of the $d_{xy}$ orbital 
to the $d_{x^{2}\textrm{-}y^{2}}$ orbital; 
instead, we take account of the effect of this hybridization 
as the difference of the CEF energies 
between the $d_{xz/yz}$ and $d_{xy}$ orbitals, as denoted in Sec. II. 
In contrast to this density-functional calculaiton~\cite{hyb-t2g-eg}, 
another density-functional calculation 
for $x=0.5$ within local spin-density approximation has found 
a disappearance of this nesting vector 
in the presence of the spin-orbit interaction, 
which is equal to $0.167$ eV.~\cite{Oguchi-LS} 
It is thus necessary to study 
the effects of this new FS on the electronic structures around $x=0.5$ 
in more detail. 
However, we think that 
the presence of this new FS will not lead to 
the drastic changes from the present results 
since this rotation-induced FS is small 
and the DOS for each Ru $t_{2g}$ orbital obtained in this study 
is almost the same as that obtained in these density-functional 
studies (Refs. \onlinecite{Oguchi-LS} and \onlinecite{hyb-t2g-eg}). 
 
Finally, 
we compare the obtained results with the previous study~\cite{NA-GA} 
in the Gutzwiller approximation. 
This previous study 
calculated the total renormalization factor of the kinetic energy 
for the Ru $t_{2g}$ orbitals, 
which is inversely proportional to the mass enhancement, 
and found that 
the inverse of the total renormalization factor 
is largest for the model of $x=0.5$. 
Namely, these results can reproduce the experimentally observed tendency of 
the effective mass in $0.5\leq x \leq 2$ (Ref. \onlinecite{Nakatsuji-HF}). 
These results suggest that 
the criticality approaching the usual Mott transition plays 
an important role in enhancing the effective mass towards $x=0.5$; 
in this usual Mott insulator, 
the occupation numbers for the $d_{xz/yz}$ and $d_{xy}$ orbitals 
are 1 and 2, respectively. 
On the other hand, 
in this study, we find that 
several modes of spin fluctuation for the $d_{xy}$ orbital 
around $\boldq=(0,0)$ and $\boldq\sim (0.797\pi,0)$ 
are strongly enhanced for the model of $x=0.5$. 
As discussed above, 
this enhancement of several modes will lead to 
the larger mass enhancement for $x=0.5$ than for $x=2$. 
This mechanism of the mass enhancement can coexist with the proposal 
in the previous study~\cite{NA-GA} in the Gutzwiller approximation 
since the former and latter mass enhancements arise 
from the nonlocal and local effects. 
In other words, 
not only the criticality approaching the usual Mott transition, 
but also the spin fluctuations for the $d_{xy}$ orbital 
around $\boldq=(0,0)$ and $\boldq\sim (\pi,0)$ 
play an important role in enhancing the effective mass towards $x=0.5$. 

\subsection{Comparison with experimental results}
We first address the correspondence of our results 
with several experimental results at $x=2$. 
In addition to the inelastic neutron measurement,~\cite{Neutron-Sr214} 
our results are consistent with 
the nuclear magnetic resonance (NMR) measurement.~\cite{Ishida-NMR} 
This experiment has found that 
not only the IC AF spin fluctuation at $\boldq\sim(0.6\pi,0.6\pi)$ 
but also the FM spin fluctuation plays 
the non-negligible role in enhancing the nuclear spin-lattice relaxation rate 
at a Ru site. 
This result indicates that 
the FM spin fluctuation also plays 
an important role in discussing the magnetic properties for $x=2$. 
Actually, 
we see that 
the spin fluctuation around $\boldq=(0,0)$ is non-negligible, 
although the dominant components arise from 
the IC spin fluctuations around $\boldq\sim (\pi,0.5\pi)$, 
$(0.235\pi,0.235\pi)$, $(\pi,0.656\pi)$, and $(0.688\pi,0.688\pi)$. 

Next, we compare our results with several experimental results 
in the presence of the rotation of RuO$_{6}$ octahedra. 
A recent muon spin relaxation ($\mu$SR) measurement~\cite{Wrong-muSR} 
for Ca$_{2-x}$Sr$_{x}$RuO$_{4}$ with $x=0.5$ and $1.5$ 
and Sr$_{2}$Ru$_{1-y}$Ti$_{y}$O$_{4}$ with $y=0.09$ 
has shown that 
the muon relaxation rates rapidly increase as temperature decreases, 
and that the zero-field $\mu$SR time spectra  
are similar to those for dilute-alloy spin glasses. 
The authors have proposed the static spin glass order 
not only in Sr$_{2}$Ru$_{1-y}$Ti$_{y}$O$_{4}$ 
but also in Ca$_{2-x}$Sr$_{x}$RuO$_{4}$ for $0.2\leq x \leq 1.6$. 
However, 
all the previous experimental studies for Ca$_{2-x}$Sr$_{x}$RuO$_{4}$ 
in $0.5\leq x \leq 2$ (Refs.~\onlinecite{Nakatsuji-discovery,Nakatsuji-lattice,
Nakatsuji-HF,Neutron-gamma,Neutron-New,Ishida-NMR-x05,Hall-CSRO,xray-CSRO}) 
have suggested that 
the ground states are the PM metals. 
In addition, the elastic neutron measurement for $x=1.5$ by the authors 
for this $\mu$SR measurement has observed the absence of the ordered moment 
for $x=1.5$, 
while the $\mu$SR measurement has estimated the ordered moment 
at $0.25\mu_{\textrm{B}}$ per Ru. 
Therefore, we think that 
the conclusion for this $\mu$SR measurement is doubtful, 
and that 
it is reasonable to analyze the static susceptibilities in the PM states 
to discuss the electronic states around $x=0.5$. 

We now discuss the correspondence with several neutron measurements. 
The polarized neutron diffraction measurement~\cite{polarized-neutron} 
for $x=0.5$ 
has observed the anisotropic spin density distribution at a Ru site, 
which is flattened along $c$ axis. 
This result suggests that 
the main part of the spin density arises from that for the $d_{xy}$ orbital. 
Namely, the spin fluctuation for the $d_{xy}$ orbital is predominant at $x=0.5$. 
Actually, 
we find that the dominant contributions to the spin fluctuation 
arise from that for the $d_{xy}$ orbital, 
which are consistent with this experimental result. 

In addition to this neutron measurement, 
the inelastic neutron measurements~\cite{Neutron-gamma,Neutron-New} 
for $x=0.62$ have found that 
the FM scattering is maximum at an energy transfer of $0.4$ meV, 
while the IC scatterings with $\boldq=(0.12,0,0)$ and $(0.27,0,0)$ 
evolve at an energy transfer of $\sim2.5$ meV. 
These results indicate that 
the FM spin fluctuation is predominant at a lower energy transfer, 
while the IC spin fluctuations evolve as an energy transfer increases. 
Furthermore, these measurements~\cite{Neutron-gamma,Neutron-New} have observed 
the intensity spreading in momentum space around the maximum peak 
and the large amplitude of the real part of the dynamic susceptibility 
at the maximum peak. 
These results indicate that 
several modes of spin fluctuation are more strongly enhanced for $x=0.62$ 
than for $x=2$, 
and that 
Ca$_{2-x}$Sr$_{x}$RuO$_{4}$ with $x=0.62$ is located nearer to 
the boundary of the magnetic instability. 
Therefore, 
all the results for the model of $x=0.5$ and the special model 
are qualitatively consistent with these experimental results 
except the evolution of the IC scatterings. 
It is necessary to study 
the origin of the evolution of these IC scatterings. 

Our result for the model of $x=0.5$ is also consistent with 
the NMR measuerement,~\cite{Ishida-NMR-x05}   
in which the evolution of the FM spin fluctuation has been observed 
towards $x=0.5$. 
This result of the NMR measurement is consistent with 
the dependence of the Wilson ratio on the Sr concentration.~\cite{Nakatsuji-HF} 

We now briefly remark on the metamagnetic transition 
observed for Ca$_{2-x}$Sr$_{x}$RuO$_{4}$ 
in the range of $0.2\leq x \leq 0.5$.~\cite{Nakatsuji-Metamag,Nakatsuji-HF} 
In this study, 
we focus on the electronic states for $0.5\leq x \leq 2$ 
only in the absence of the external magnetic field. 
It is necessary to study 
the effects of the external magnetic field on the 
fluctuations for charge and spin sectors; 
this issue is a remaining future work. 

Finally, we remark on the Curie-Weiss (CW) behavior in the spin susceptibility 
with a Curie constant corresponding to nearly $S=1/2$ 
for $0.5\leq x < 2$.~\cite{Nakatsuji-discovery,Nakatsuji-lattice} 
The origin of this CW behavior has not been clarified yet. 
In general, 
there are two mechanisms in emerging 
the CW-type temperature dependence in the spin susceptibility.~\cite{Moriya-review,Moriya-CW} 
One is due to the formation of the localized moment, 
and the other is due to the drastic renormalization of 
the temperature dependence for the free energy by enhanced fluctuations. 
In the latter mechanism, 
the mode-mode coupling 
leads to this drastic renormalization.~\cite{Moriya-SCR,Moriya-SCR-AF} 
The emergence of the CW behavior in Ca$_{2-x}$Sr$_{x}$RuO$_{4}$ 
will arise from the latter mechanism 
since all electrons for the Ru $t_{2g}$ orbitals are itinerant 
in $0.5\leq x \leq 2$.~\cite{ARPES05,opt-mass} 
In addition, 
we find for the model of $x=0.5$ that 
several modes of spin fluctuation for the $d_{xy}$ orbital 
around $\boldq=(0,0)$ and $\boldq\sim (0.797\pi,0)$ 
are strongly enhanced, 
and that the dominant contributions to the susceptibilities 
for a spin sector arise from those for the $d_{xy}$ orbital. 
Therefore, 
we think that 
the mode-mode coupling for the $d_{xy}$ orbital 
will play a more important role 
in discussing the spin susceptibility around $x=0.5$ 
than for the $d_{xz/yz}$ orbital. 
It is necessary to study the electronic states around $x=0.5$ 
on the basis of the microscopic theory taking account of 
the mode-mode coupling. 

\section{Summary}
To clarify the effects of the nonlocal (but not long-range) correlation 
on the electronic structures in Ca$_{2-x}$Sr$_{x}$RuO$_{4}$ around $x=0.5$, 
we studied the static susceptibilities for charge and spin sectors 
in the PM states within the RPA 
for the models of $x=2$ and $0.5$, and a special model. 
In particular, 
we analyzed the effects of the rotation of RuO$_{6}$ octahedra, 
the vHs for the $d_{xy}$ orbital, and the Hund's rule coupling 
on these static susceptibilities. 

First, we analyzed the effects of the rotation of RuO$_{6}$ octahedra 
by comparing the results at $J_{\textrm{H}}=U/6$ 
for the models of $x=2$ and $0.5$. 
In the absence of interactions, 
we found that 
the rotation of RuO$_{6}$ octahedra leads to 
the increase of the noninteracting susceptibility 
for the $d_{xy}$ orbital around $\boldq=(0,0)$ and 
the shifts of the IC peak for the $d_{xz/yz}$ and $d_{xy}$ orbitals 
towards $\boldq=(\pi,\pi)$ and $(\pi,0)$. 
For the model of $x=2$ in the RPA, 
we found that 
$\lambda^{\textrm{S}}_{\textrm{max}}(\boldq)^{-1}$ at $\boldq\sim (\pi,0.5\pi)$ 
touches zero at $U=U_{\textrm{c}}=0.975$, 
and that 
there are two main peaks in $\lambda^{\textrm{S}}_{\textrm{max}}(\boldq)^{-1}$ 
at $\boldq\sim(\pi,0.5\pi)$ and $(0.235\pi,0.235\pi)$ 
and two secondary peaks at $\boldq\sim(\pi,0.656\pi)$ 
and $(0.688\pi,0.688\pi)$. 
The two main peaks arise from 
the corresponding fluctuations for the $d_{xy}$ orbital, 
and the secondary peaks at $\boldq\sim(\pi,0.656\pi)$ 
and $(0.688\pi,0.688\pi)$ arise from 
the corresponding fluctuation for the $d_{yz}$ orbital and 
the combined fluctuation of the $d_{xz/yz}$ and $d_{xy}$ orbitals. 
We also found that 
$\chi^{\textrm{S}}(\boldq,0)$ with $\boldq\sim (\pi,0.5\pi)$, 
$(\pi,0.656\pi)$, and $(0.688\pi,0.688\pi)$ 
are dominant at $U=0.8U_{\textrm{c}}$, 
while $\chi^{\textrm{S}}(\boldq,0)$ with $\boldq\sim(0.688\pi,0.688\pi)$ 
becomes largest at $U=0.4U_{\textrm{c}}$. 
The main peak in $\lambda^{\textrm{S}}_{\textrm{max}}(\boldq)^{-1}$ 
at $\boldq\sim(0.235\pi,0.235\pi)$ is less important at 
$U=0.8U_{\textrm{c}}$ and $0.4U_{\textrm{c}}$. 
For the model of $x=0.5$ in the RPA, 
we found that 
$\lambda^{\textrm{S}}_{\textrm{max}}(\boldq)^{-1}$ at $\boldq=(0,0)$ 
touches zero at $U=U_{\textrm{c}}=0.751$, which 
is smaller than for the model of $x=2$, 
and that 
there are two main peaks in $\lambda^{\textrm{S}}_{\textrm{max}}(\boldq)^{-1}$ 
at $\boldq=(0,0)$ and $\boldq\sim(0.797\pi,0)$ 
and a secondary peak at $\boldq\sim(\pi,0.125\pi)$. 
In contrast to the case of $x=2$, 
all the peaks in $\lambda^{\textrm{S}}_{\textrm{max}}(\boldq)^{-1}$ 
arise from the corresponding susceptibilities for the $d_{xy}$ orbital. 
We also found that 
$\chi^{\textrm{S}}(\boldq,0)$ at $\boldq=(0,0)$ 
and $\boldq\sim(\pi,0.125\pi)$ 
are dominant at $U=0.8U_{\textrm{c}}$, 
while $\chi^{\textrm{S}}(\boldq,0)$ at $\boldq=(0,0)$, 
$\boldq\sim(0.797\pi,0)$, and $(\pi,0.125\pi)$ 
are nearly same at $U=0.4U_{\textrm{c}}$. 
The values of $\chi^{\textrm{S}}(\boldq,0)$ 
along $(0,0)\rightarrow (\pi,0)$ 
are larger than along $(0,0)\rightarrow (\pi/2,\pi/2)$. 
Furthermore, 
we found that 
the peaks in $\lambda^{\textrm{S}}_{\textrm{max}}(\boldq)^{-1}$ 
for the model of $x=0.5$ are less sharp than for the model of $x=2$. 
These less sharp peaks and the smaller value of $U_{\textrm{c}}$ 
will lead to a larger effective mass for the model of $x=0.5$ 
than for the model of $x=2$. 

Second, to analyze the effects of the vHs for the $d_{xy}$ orbital, 
we compared the results at $J_{\textrm{H}}=U/6$ for the special model 
with those for the model of $x=0.5$. 
In the absence of interactions, 
we found the larger increase of the noninteracting susceptibility 
for the $d_{xy}$ orbital and the smaller shifts of the IC peak 
for the $d_{xz/yz}$ and $d_{xy}$ orbitals towards $\boldq=(\pi,\pi)$ 
and $(\pi,0)$ in the special model. 
The former is due to the larger increase of the DOS for the $d_{xy}$ orbital 
near the Fermi level, 
and the latter is due to the smaller value of $\Delta_{t_{2g}}$. 
In the presence of interactions, 
we found that 
the value of $U_{\textrm{c}}$, 
where $\lambda^{\textrm{S}}_{\textrm{max}}(\boldq)^{-1}$ 
at $\boldq=(0,0)$ touches zero, 
is smaller for the special model than for the model of $x=0.5$, 
and that 
there is a main peak at $\boldq=(0,0)$ and there is a secondary peak 
at $\boldq\sim(0.703\pi,0)$; 
the latter peak corresponds to the peak at $\boldq\sim(0.797\pi,0)$ 
for the model of $x=0.5$. 
We also found that 
all the peaks in $\lambda^{\textrm{S}}_{\textrm{max}}(\boldq)^{-1}$ 
for the special model are slightly sharper than for the model of $x=0.5$, 
and that 
the competition between the modes of 
spin fluctuation around $\boldq=(0,0)$ and $\boldq\sim (\pi,0)$ 
is weaker for the special model. 
Furthermore, we found that 
all the enhanced modes in $\lambda^{\textrm{S}}_{\textrm{max}}(\boldq)^{-1}$ 
arise from the corresponding fluctuations for the $d_{xy}$ orbital, 
and that 
the dominant contributions to the static susceptibility 
for a spin sector arise from those for the $d_{xy}$ orbital 
around $\boldq=(0,0)$ and $\boldq\sim (\pi,0)$; 
these results are qualitatively the same as those for the model of $x=0.5$. 

Third, we analyzed the $J_{\textrm{H}}$ dependencies of the static susceptibilities 
for the models of $x=2$ and $0.5$, and the special model. 
For all these models, 
we found that 
the dominant wave vectors of the static susceptibilities 
at $J_{\textrm{H}}=0$ and $U/4$ are the same as those at $J_{\textrm{H}}=U/6$, 
that the increase of $J_{\textrm{H}}/U$ leads to 
the decrease of $U_{\textrm{c}}$, 
and that the static susceptibility for a charge sector 
is enhanced only at $J_{\textrm{H}}=0$. 
In addition to these, 
we found for the model of $x=2$ that 
the increase of $J_{\textrm{H}}/U$ enhances 
the IC AF spin fluctuations at $\boldq\sim (0.688\pi,0.688\pi)$ 
and $(\pi,0.656\pi)$, 
and that the former IC AF spin fluctuation 
is strongest for $J_{\textrm{H}}\geq U/6$ at $U=0.8U_{\textrm{c}}\sim 0.78$; 
these modes correspond to the secondary peaks 
in $\lambda^{\textrm{S}}_{\textrm{max}}(\boldq)^{-1}$ at $J_{\textrm{H}}=U/6$. 
The enhancement of the IC AF spin fluctuation 
at $\boldq\sim(0.688\pi,0.688\pi)$ 
arises not only from the $d_{xz/yz}$ orbital but also from the $d_{xy}$ orbital. 
Although the similar results have been obtained 
in the previous mean-field study~\cite{Nomura-MFA} for $x=2$, 
we think that a set of the parameters used in this study is more realistic. 
In contrast to the case of $x=2$, 
the increase of $J_{\textrm{H}}/U$ leads to a nearly uniform enhancement 
of $\chi^{\textrm{S}}(\boldq,0)$ for the model of $x=0.5$ and the special model. 
This difference is due to the much larger contribution from the $d_{xy}$ orbital 
than from the $d_{xz/yz}$ orbital 
in the presence of the rotation of RuO$_{6}$ octahedra. 

In summary, the analyses in the RPA reveal that 
the rotation of RuO$_{6}$ octahedra leads to 
the enhancement of several modes of spin fluctuation for the $d_{xy}$ orbital 
around $\boldq=(0,0)$ and $\boldq\sim (\pi,0)$. 
This enhancement arises from 
the increase of the corresponding susceptibilities 
for the $d_{xy}$ orbital due to 
the rotation-induced modifications of the electronic structure 
for this orbital (i.e., the flattening of the bandwidth 
and the increase of the DOS near the Fermi level). 
These analyses also reveal that 
the location of the vHs for the $d_{xy}$ orbital, 
which is controlled by substitution of Ca for Sr, 
is a parameter to control the competition between 
the modes of spin fluctuation for the $d_{xy}$ orbital 
around $\boldq=(0,0)$ and $\boldq\sim (\pi,0)$. 
We propose that 
the spin fluctuations for the $d_{xy}$ orbital 
around $\boldq=(0,0)$ and $\boldq\sim (\pi,0)$ 
play an important role in the electronic states around $x=0.5$ 
other than the criticality approaching the usual Mott transition 
where all electrons are localized.

\begin{acknowledgments}
The authors would like to thank Y. Yanase for a useful comment 
about the obtained results 
and T. Kariyado for useful comments about the numerical calculations. 
\end{acknowledgments}


\end{document}